\documentclass[aps,preprint,nofootinbib]{revtex4}
\usepackage{graphicx}
\begin{document}
\newcommand{\met}{\not\!\! E_{T}}

\begin{flushright}
UCRHEP-T437\\
 August 2007
\par\end{flushright}

\vskip 0.5in

\title{{\Large Observing the Dark Scalar Doublet and its Impact}{\normalsize{}
}\\
{\normalsize{} }{\Large on the Standard-Model Higgs Boson at Colliders
\\[50pt]}}

\author{Qing-Hong Cao}

\email{qcao@ucr.edu}

\affiliation{Department of Physics and Astronomy, University of California, Riverside,
California 92521, USA}

\author{Ernest Ma}

\email{ernest.ma@ucr.edu}

\affiliation{Department of Physics and Astronomy, University of California, Riverside,
California 92521, USA}

\author{G. Rajasekaran}

\email{graj@imsc.res.in}

\affiliation{Institute of Mathematical Sciences, Chennai (Madras) 600113, India\\
 $~$\\
 }

\begin{abstract}
If the Standard Model (SM) of particle interactions is extended to
include a second scalar doublet $[H^{+},(H^{0}+iA^{0})/\sqrt{2}]$,
which is odd under an unbroken $Z_{2}$ discrete symmetry, it may
be called the $dark$ scalar doublet, because its lightest neutral
member, say $H^{0}$, is one posssible component for the dark matter
of the Universe. We discuss the general phenomenology of the four
particles of this doublet, without assuming that $H^{0}$ is the dominant
source of dark matter. We also consider the impact of this $dark$
scalar doublet on the phenomenology of the SM Higgs boson $h$. 
\end{abstract}
\maketitle

\section{Introduction}

The canonical Standard Model (SM) of quarks and leptons requires only
one Higgs doublet $(\phi^{+},\phi^{0})$, such that $\phi^{0}$ acquires
a nonzero vacuum expectation value $\langle\phi^{0}\rangle=v=174$
GeV, from which all particles (except neutrinos) obtain mass. Suppose
a second scalar doublet $[H^{+},(H^{0}+iA^{0})/\sqrt{2}]$ is added\ \citep{Deshpande:1977rw},
which is $odd$ under an $unbroken$ $Z_{2}$ discrete symmetry, then
there are at least two interesting outcomes.

(I) This scalar doublet also has a conserved additive quantum number,
so that $(H^{0}+iA^{0})/\sqrt{2}$ is a mass eigenstate, i.e. $m_{H^{0}}=m_{A^{0}}$.
Such is in fact the case of supersymmetry, where this is nothing but
the complex conjugate of a scalar lepton doublet. More generally,
it may belong to a new class of particles\ \citep{Ma:1977dm,Ma:1977pc}
with a separately conserved additive quantum number unrelated to lepton
number.

(II) There is no additional symmetry beyond the postulated $Z_{2}$,
in which case $m_{H^{0}}\neq m_{A^{0}}$. As pointed out\ \citep{Deshpande:1977rw}
already in 1977, either $H^{0}$ or $A^{0}$ must then be stable.
This idea has been revived recently, first in the context of radiative
neutrino masses\ \citep{Ma:2006km}, then as a means of understanding
precision electroweak data\ \citep{Barbieri:2006dq}. In any case,
this specific potential new source of dark matter of the Universe\ \citep{Bertone:2004pz}
is gaining more attention\ \citep{Ma:2006fn,Majumdar:2006nt,Honorez:2006gr,Sahu:2007uh,Gustafsson:2007pc,Lisanti:2007ec}.
It will also have an important bearing on interpreting forthcoming
data from the LHC (Large Hadron Collider) at CERN.

In this paper we will analyze in detail the two-scalar-doublet structure
of scenario (II), which we call DSDM (Dark Scalar Doublet Model).
Whereas there are specific constraints from identifying $H^{0}$ as
the $sole$ source of dark matter\ \citep{Barbieri:2006dq,Honorez:2006gr},
we will take the viewpoint that there may also be other sources of
dark matter such as that coming from supersymmetry. For example, the
first such explicit model with two (and possibly three) different
dark-matter particles has already been proposed\ \citep{Ma:2006uv}.
Thus our analysis will allow a wider range of parameter values, with
the aim of extracting the possible signals of this $dark$ scalar
doublet at the LHC. This doublet has been called the {}``inert Higgs
doublet''\ \citep{Barbieri:2006dq,Honorez:2006gr}, but it is neither
inert (since it has gauge and scalar interactions) nor a contributor
to the Higgs mechanism (since it has no vacuum expectation value),
hence we prefer to call it the $dark$ scalar doublet.

Consider the scalar potential of $\Phi_{1}=(\phi^{+},\phi^{0})$ and
$\Phi_{2}=[H^{+},(H^{0}+iA^{0})/\sqrt{2}]$, with $\Phi_{1}$ even
and $\Phi_{2}$ odd under an exactly conserved $Z_{2}$: \begin{eqnarray}
V & = & \mu_{1}^{2}\Phi_{1}^{\dagger}\Phi_{1}+\mu_{2}^{2}\Phi_{2}^{\dagger}\Phi_{2}+\frac{1}{2}\lambda_{1}(\Phi_{1}^{\dagger}\Phi_{1})^{2}+\frac{1}{2}\lambda_{2}(\Phi_{2}^{\dagger}\Phi_{2})^{2}+\lambda_{3}(\Phi_{1}^{\dagger}\Phi_{1})(\Phi_{2}^{\dagger}\Phi_{2})\nonumber \\
 & + & \lambda_{4}(\Phi_{1}^{\dagger}\Phi_{2})(\Phi_{2}^{\dagger}\Phi_{1})+\frac{1}{2}\lambda_{5}(\Phi_{1}^{\dagger}\Phi_{2})^{2}+\frac{1}{2}\lambda_{5}^{*}(\Phi_{2}^{\dagger}\Phi_{1})^{2}.\end{eqnarray}
 All the above parameters are necessarily real except $\lambda_{5}$,
but the exact $Z_{2}$ symmetry (which forbids the bilinear term $\Phi_{1}^{\dagger}\Phi_{2}$
allows us to rotate the relative phase between $\Phi_{1}$ and $\Phi_{2}$
to make $\lambda_{5}$ real as well. As the standard $SU(2)\times U(1)$
gauge symmetry is spontaneously broken by $\langle\phi^{0}\rangle=v=174$
GeV, the masses of the resulting scalar particles are given by\ \citep{Ma:2006km}
\begin{eqnarray}
m^{2}(h) & = & 2\lambda_{1}v^{2},\label{eq:mass-sm-higgs}\\
m^{2}(H^{\pm}) & = & \mu_{2}^{2}+\lambda_{3}v^{2},\label{eq:mass-HP}\\
m^{2}(H^{0}) & = & \mu_{2}^{2}+(\lambda_{3}+\lambda_{4}+\lambda_{5})v^{2},\label{eq:mass-H0}\\
m^{2}(A^{0}) & = & \mu_{2}^{2}+(\lambda_{3}+\lambda_{4}-\lambda_{5})v^{2}.\label{eq:mass-A0}\end{eqnarray}
 The lone Higgs boson of the Standard Model is of course $h$, whereas
$H^{\pm}$, $H^{0}$, and $A^{0}$ are the components of the $dark$
scalar doublet which interact with $h$ and among themselves as follows:
\begin{eqnarray}
V_{int} & = & \frac{1}{2}\lambda_{2}\left[H^{+}H^{-}+\frac{1}{2}(H^{0})^{2}+\frac{1}{2}(A^{0})^{2}\right]^{2}+\lambda_{3}\left(vh+\frac{1}{2}h^{2}\right)\left[H^{+}H^{-}+\frac{1}{2}(H^{0})^{2}+\frac{1}{2}(A^{0})^{2}\right]\nonumber \\
 & + & \frac{1}{2}(\lambda_{4}+\lambda_{5})\left(vh+\frac{1}{2}h^{2}\right)(H^{0})^{2}+\frac{1}{2}(\lambda_{4}-\lambda_{5})\left(vh+\frac{1}{2}h^{2}\right)(A^{0})^{2}.\end{eqnarray}
 Being electroweak doublets, they also have gauge interactions, but
they do not interact directly with quarks or leptons.

Assuming the mass hierarchy $m_{H^{\pm}}>m_{A^{0}}>m_{H^{0}}$, the
stable $H^{0}$ then appears as missing energy in the decays of $A^{0}$
and $H^{\pm}$. Since there is no term linear in $A^{0}$ or $H^{0}$
in Eq.~(6), the decay of $A^{0}$ must occur through the gauge interaction
\begin{equation}
\frac{g}{2\cos\theta_{W}}Z_{\mu}(H^{0}\partial^{\mu}A^{0}-A^{0}\partial^{\mu}H^{0}).\end{equation}
 Hence the dominant decay of $A^{0}$ is into $H^{0}f\bar{f}$, where
$f$ = lepton or quark, and $H^{0}$ is missing energy.

As for $H^{\pm}$, its gauge interactions with $H^{0}$ and $A^{0}$
are given by \begin{equation}
\frac{ig}{2}W_{\mu}^{-}(H^{0}\partial^{\mu}H^{+}-H^{+}\partial^{\mu}H^{0})+\frac{g}{2}W_{\mu}^{-}(A^{0}\partial^{\mu}H^{+}-H^{+}\partial^{\mu}A^{0})+h.c.\end{equation}
 Hence the dominant decays of $H^{\pm}$ are into $W^{\pm}H^{0}$
and $W^{\pm}A^{0}$, with the subsequent decay of $A^{0}$ into $Z^{0}H^{0}$
in the latter.

The only other trilinear gauge interactions of the $dark$ scalars
are those of $H^{\pm}$ with $Z$ and $\gamma$. There are of course
also the quadrilinear terms required by gauge invariance.

\section{LEP constraints}

\subsection{$Z$-boson decay}

If $m_{H^{0}}+m_{A^{0}}<m_{Z}$, then $Z$ can decay into $H^{0}A^{0}$
with the subsequent decay of $A^{0}$ to $H^{0}f\bar{f}$ where $f$
is a quark or lepton. This would appear as a rare $Z$ decay event
with $f\bar{f}$ + missing energy in the final state. The high precision
LEP-I experiments have accumulated a sample of about $1.7\times10^{7}$
on-shell $Z$ boson decays via the $q\bar{q}$ and $\ell^{+}\ell^{-}$
channels\ \citep{LEPII:2005em}. ALEPH preformed a partial analysis
of an integrated $79\,{\rm pb}^{-1}$ to search for the rare decay
$Z\to f\bar{f}+\nu\bar{\nu}$ and found no event above the SM prediction\ \citep{Kobel:1996re}.
To estimate how the whole LEP-I data sample can constrain the DSDM,
we calculate the SM partial decay width of $Z\to f\bar{f}+\nu\bar{\nu}$
($f=q,\ell$) and obtain its decay branching ratio \begin{eqnarray*}
{\rm BR}\left(Z\to f\bar{f}+\nu\bar{\nu}\right) & = & \frac{\left(4.269_{q\bar{q}\nu\bar{\nu}}+0.779_{\ell^{+}\ell^{-}\nu\bar{\nu}}\right)\times10^{-7}\,{\rm GeV}}{2.494_{{\rm total}}\,{\rm GeV}}\\
 & \simeq & 2.02\times10^{-7}.\end{eqnarray*}
 We thus expect about $\left(1.7\times10^{7}\right)\times\left(2.0\times10^{-7}\right)\simeq3.4$
background events from the SM. Assuming that only 4 events show up
as $f\bar{f}$ + missing energy from the whole LEP-I data sample,
we deduce according to Possion statistics that only about 5 signal
events are allowed at $95\%$ C.L. in the $f\bar{f}+H^{0}H^{0}$ channel,
i.e. \[
\frac{\Gamma\left(Z\to f\bar{f}+H^{0}H^{0}\right)}{\Gamma_{0}}\leq\frac{5}{1.7\times10^{7}\times0.8}\simeq3.7\times10^{-7},\]
 where the SM branching ratio $BR\left(Z\to f\bar{f}\right)\simeq0.8$
(for $f=q,\ell$) has been used. This implies that the phase space
for $Z\to H^{0}A^{0}$ is very small (with a mass gap of a few MeV),
so for all practical purposes, the bound $m_{H^{0}}+m_{A^{0}}>m_{Z}$
should be respected\ \citep{Gustafsson:2007pc}. The resulting lower
limits on $m_{H^{0}}$ for various $\Delta m_{A^{0}H^{0}}\equiv m_{A^{0}}-m_{H^{0}}$
are shown in Table\ \ref{tab:zpole}.

\begin{table}[htb]
\caption{Lower limits on $m_{H^{0}}$ from the LEP $Z$-decay constraint for
various $\Delta m_{A^{0}H^{0}}$.\label{tab:zpole}}

\begin{tabular}{ccccccccccc}
\hline 
$\Delta m_{A^{0}H^{0}}\,\left({\rm GeV}\right)$ & \ \ \ \  & $5$ & \ \ \ \  & 10 & \ \ \ \  & 15 & \ \ \ \  & 20  & \ \ \ \  & 30\tabularnewline
\hline
\hline 
$m_{H^{0}}\,\left({\rm GeV}\right)$ &  & 43 &  & 41 &  & 38 &  & 36 &  & 31\tabularnewline
\hline
\end{tabular}
\end{table}

Since $A^{0}\to H^{0}+\nu\bar{\nu}$ is also possible, this process
may also contribute to the invisible width of the $Z$ if $m_{H^{0}}+m_{A^{0}}<m_{Z}$.
However, the corresponding experimental uncertainty allows for a mass
gap of a few GeV, which is not competitive with the constraint discussed
above.

\subsection{Neutralino search at LEP}

While there is no existing experimental search that can be directly
translated into a search for the dark scalars, the search for charginos
and neutralinos is very close as they may exhibit the same collider
signatures. The OPAL Collaboration\ \citep{Abbiendi:2003sc} has
performed a search for neutralino pair production via the process
of $e^{+}e^{-}\to\widetilde{\chi}_{2}^{0}\widetilde{\chi}_{1}^{0}$
in the hadronic decay mode $\widetilde{\chi}_{2}^{0}\to\widetilde{\chi}_{1}^{0}Z^{*}\to\widetilde{\chi}_{1}^{0}qq$
where $\widetilde{\chi}_{1,2}^{0}$ denotes the lightest (next lightest)
neutralino, respectively. No significant excess above the SM background
leads to an upper limit on the cross section of $\widetilde{\chi}_{2}^{0}\widetilde{\chi}_{1}^{0}$
associated production: $\sigma_{\widetilde{\chi}_{1}^{0}\widetilde{\chi}_{2}^{0}}\sim0.1\,{\rm pb}-0.5\,{\rm pb}$
for $m_{\widetilde{\chi}_{1}^{0}}\sim50\,{\rm GeV}$ and $m_{\tilde{\chi}_{2}^{0}}\sim60-80\,{\rm GeV}$
(see the right figure of Fig.\ 10 in Ref.\ \citep{Abbiendi:2003sc}).
The L3 Collaboration\ \citep{Acciarri:1999km} also performed a similar
search at $\sqrt{s}=189\,{\rm GeV}$, which gives the upper limit
on the cross section as $\sigma_{\widetilde{\chi}_{1}^{0}\widetilde{\chi}_{2}^{0}}\sim0.1\,{\rm pb}-2\,{\rm pb}$
for $m_{\widetilde{\chi}_{1}^{0}}\sim50\,{\rm GeV}$ and $m_{\tilde{\chi}_{2}^{0}}\sim60-80\,{\rm GeV}$
(see Fig.\ 3a in Ref.\ \citep{Acciarri:1999km}).

The translation of the above limits on the cross section from the
neutralino to the dark scalar is not straightforward even though they
share the same topologies. The kinematics and spin correlations of
$\widetilde{\chi}_{1}^{0}\widetilde{\chi}_{2}^{0}$ and $A^{0}H^{0}$
pair productions are totally different. For example, the scalar pair
has to be in the $p$-wave as it is produced via a virtual $Z$-boson,
therefore, each scalar has large transverse momentum\ \citep{Cao:2003tr}.
Furthermore, there is no spin correlation between the two scalars.
Those differences lead to different acceptances of the kinematics
cuts used by the OPAL and L3 Collaborations to extract the signal
from the background. One thus cannot simply apply the OPAL and L3
constraints to the DSDM. In order to get the realistic limits, one
has to go through the whole analysis using the correct kinematics
and spin correlations. For a rough estimation, we now simply compare
the prediction of the DSDM to the OPAL and L3 limits. The production
cross section of $e^{+}e^{-}\to A^{0}H^{0}$ in the DSDM at $\sqrt{s}=208\,{\rm GeV}$
is about $0.28\,{\rm pb}-0.20\,{\rm pb}$ for $m_{H^{0}}=50\,{\rm GeV}$
and $m_{A^{0}}=60\,{\rm GeV}-80\,{\rm GeV}$. It is reasonably consistent
with both OPAL and L3 constraints. Hence, we will use these masses
in the following monte carlo study.

\subsection{Higgs direct search}

\begin{figure}
\includegraphics[scale=0.6]{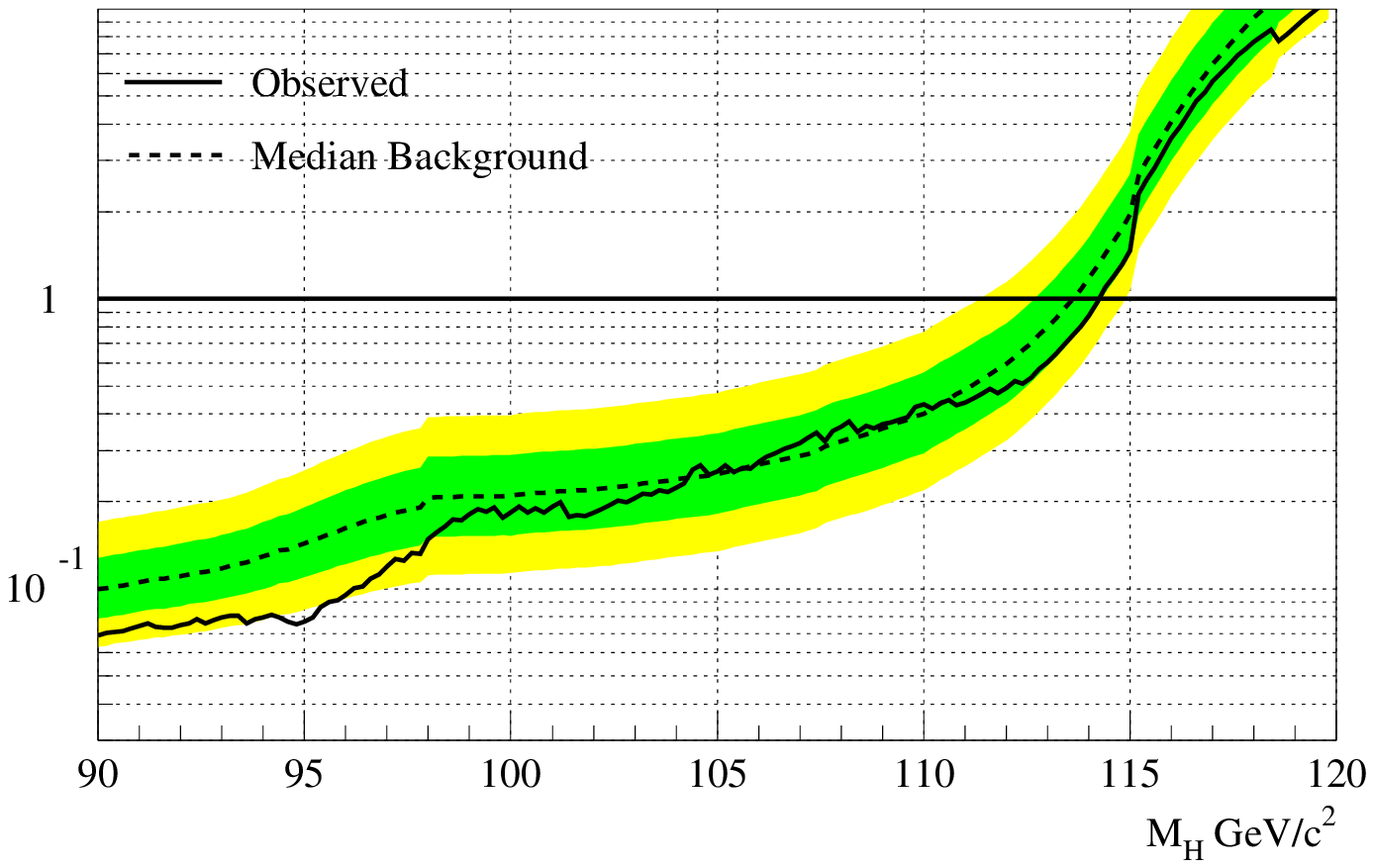}~~~~~~\includegraphics[clip,scale=0.3]{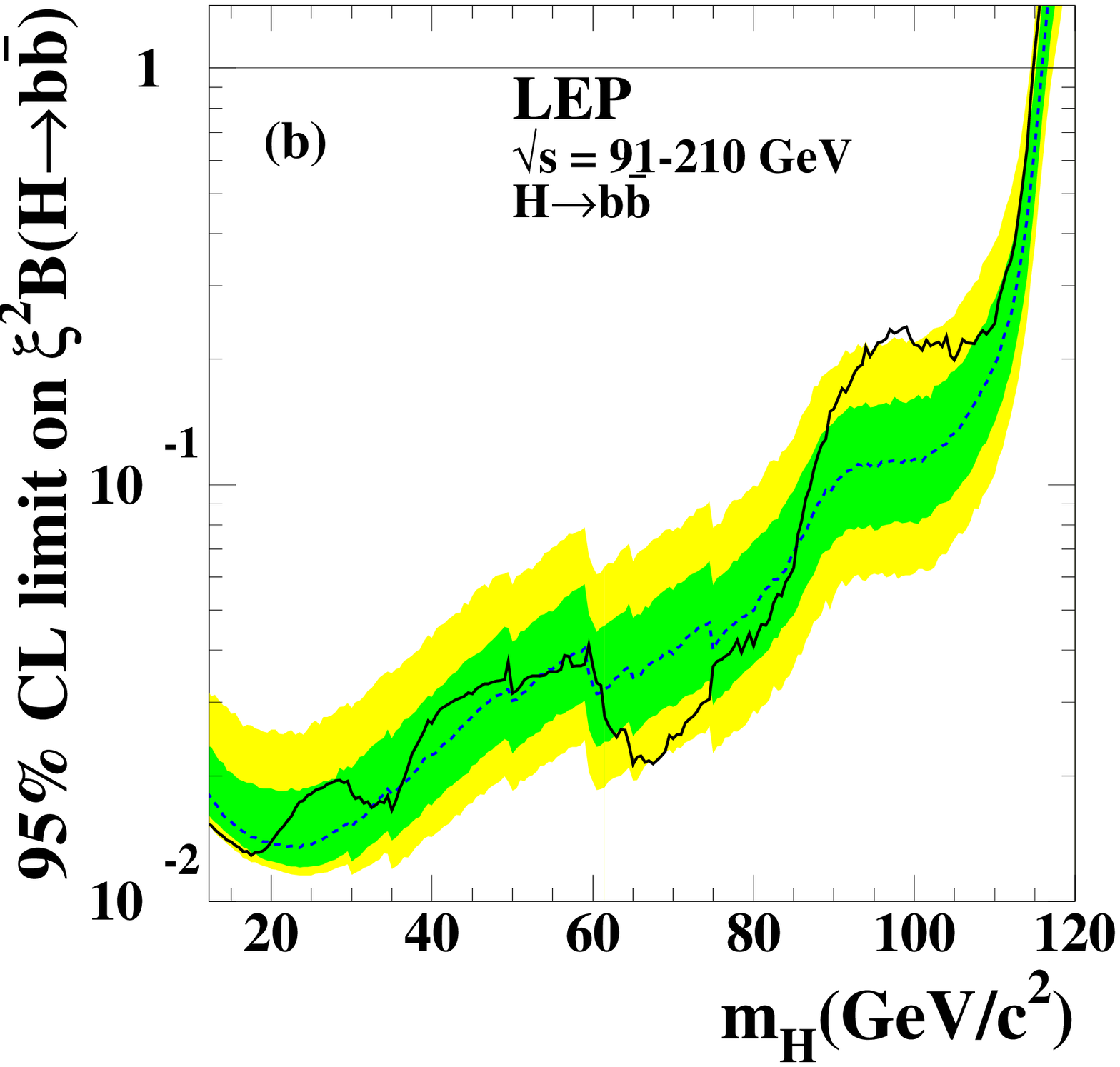}

\caption{(left) The $95\%$ C.L. upper limit on $BR\left(h\to invisible\right)$,
adapted from\ \citep{LEPII:2001xz}; (right) The $95\%$ C.L. upper
limit on $BR(h\to b\bar{b})$, adapted from\ \citep{Barate:2003sz}.
The label H in these figures refers to the SM Higgs boson $h$, and
$\xi^{2}$ in the right figure is defined as $\xi^{2}\equiv\left(g_{HZZ}/g_{HZZ}^{SM}\right)^{2}$
(see Ref.\ \citep{Barate:2003sz} for details). \label{fig:lepconstraint}}

\end{figure}

If the $Z_{2}$-odd scalars are lighter than the SM Higgs boson, they
may have a great impact on the direct search of the latter because
$h\to H^{0}H^{0}$ and $h\to A^{0}A^{0}$ may become the dominant
decay channels. In particular, the $h\to H^{0}H^{0}$ channel is invisible.
LEP II has performed a direct search of the SM Higgs boson in its
invisible decay mode but found no event, thus obtaining an upper bound
on this decay branching ratio, shown on the left in Fig.\ \ref{fig:lepconstraint}.
Below we will use this information to find the allowed parameter space
of the DSDM. As for the $h\to A^{0}A^{0}$ decay mode, it may give
rise to four charged leptons plus missing energy, resulting from $A^{0}\to H^{0}Z^{*}\to H^{0}\ell^{+}\ell^{-}$.
At LEP the SM Higgs boson is produced in association with an on-shell
$Z$-boson, i.e. $e^{+}e^{-}\to Zh$, which decays into two charged
leptons or two jets. Thus, if the SM Higgs boson decays into two $A^{0}$
bosons, there could be a collider signature of six charged leptons
plus missing energy, or four charged leptons plus two jets plus missing
energy. A new scan over the accumulated data sample at LEP to search
for such a signature may prove to be interesting.

LEP II searched the SM Higgs boson mainly via the process $e^{+}e^{-}\to Z(\to\ell^{+}\ell^{-})h(\to b\bar{b})$
as the decay $h\to b\bar{b}$ dominates for a light Higgs boson. The
negative search at LEP II gives rise to the upper bound on the decay
branching ratio of $h\to b\bar{b}$, shown in the right in Fig.\ \ref{fig:lepconstraint},
and sets a lower limit of the SM Higgs boson mass of $m_{h}>114.4\,{\rm GeV}$.
However, this limit can be relaxed in the DSDM because the invisible
decay of the Higgs boson may be dominant so that the decay branching
ratio into $b\bar{b}$ is highly suppressed\ %
\footnote{Similar studies in various models have been carried out in Refs.\ \citep{Belotsky:2002ym,Cao:2004tu,Kane:2004tk,Kim:2006mb,Belyaev:2006rf}.%
}.

The LEP II bounds on $h\to invisible$ and $h\to b\bar{b}$ compete
with each other. The former requires smaller invisible decay branching
ratio while the latter requires larger. Consider now the possibility
of the DSDM satisfying both limits.

The partial decay width of $h_{SM}\to b\bar{b}$ is given by\[
\Gamma\left(h_{SM}\to b\bar{b}\right)=\frac{N_{C}m_{b}^{2}m_{h}}{8\pi v^{2}}\left(1-\frac{4m_{b}^{2}}{m_{h}^{2}}\right)^{3/2},\]
 while the partial decay width of $h_{SM}\to SS$ ($S$ denoting a
member of the $dark$ scalar doublet) is given by \begin{equation}
\Gamma\left(h\to SS\right)=\delta_{S}\frac{v^{2}}{16\pi m_{h}}\lambda_{S}^{2}\sqrt{1-\frac{4m_{S}^{2}}{m_{h}^{2}}},\label{eq:higgs-dec-width}\end{equation}
 where $\delta_{H^{0}/A^{0}}=1$ and $\delta_{H^{+}}=2$, and the
coupling $\lambda_{S}$ is given by \begin{equation}
\lambda_{H^{0}}=\lambda_{3}+\lambda_{4}+\lambda_{5},\qquad\lambda_{A^{0}}=\lambda_{3}+\lambda_{4}-\lambda_{5},\qquad\lambda_{H^{+}}=\lambda_{3}.\label{eq:couplings}\end{equation}
 From Eqs.\ (\ref{eq:mass-HP}-\ref{eq:mass-A0}), the above couplings
can be further simplified as \begin{equation}
\lambda_{H^{0}}=\frac{m_{H^{0}}^{2}-\mu_{2}^{2}}{v^{2}},\qquad\lambda_{A^{0}}=\frac{m_{A^{0}}^{2}-\mu_{2}^{2}}{v^{2}},\qquad\lambda_{H^{+}}=\frac{m_{H^{+}}^{2}-\mu_{2}^{2}}{v^{2}}.\label{eq:couplings2}\end{equation}
 Thus, the partial decay width of the SM Higgs boson, given in Eq.\ \ref{eq:higgs-dec-width},
only depends on two parameters, $m_{S}$ and $\mu_{2}$, for a given
$m_{h}$. We note that $\mu_{2}$ is not a free parameter because
of the requirement of vacuum stability, i.e. \begin{equation}
\lambda_{1,2}>0;\qquad\lambda_{3},~\lambda_{3}+\lambda_{4}-\left|\lambda_{5}\right|>-\sqrt{\lambda_{1}\lambda_{2}}.\label{eq:vacuumstability}\end{equation}
 Hence we obtain the following bound on $\mu_{2}$, \begin{equation}
\mu_{2}^{2}<m_{H^{0}}^{2}+\sqrt{\frac{\lambda_{2}}{2}}m_{h}v,\label{eq:vacuumbound}\end{equation}
 which reveals the fact that the self-interaction of the $dark$ scalar
doublet, i.e. $\lambda_{2}$, has a nontrivial impact on the SM sector.
As to be shown later, the stronger the dark sector interacts, the
larger is the allowed parameter space of the DSDM for the SM Higgs
boson.

Assuming the charged Higgs boson mass is much larger than $m_{h}$
and $m_{A^{0}}-m_{H^{0}}=10\,{\rm GeV}$, we now scan over the parameter
space of the DSDM. In summary, the following conditions need to be
satisfied in order to relax the LEP limit $m_{h}>114.4\,{\rm GeV}$:

\begin{itemize}
\item LEP II bound, $BR_{max}\left(inv\right)$, on $h\to{invisible}$\begin{equation}
BR\left(inv\right)\equiv\frac{\Gamma\left(h_{SM}\to H^{0}H^{0}\right)}{\Gamma\left(h_{SM}\to b\bar{b}\right)+\Gamma\left(h_{SM}\to H^{0}H^{0}\right)+\Gamma\left(h_{SM}\to A^{0}A^{0}\right)}<BR_{max}\left(inv\right).\end{equation}

\item LEP II bound, $BR_{max}\left(b\bar{b}\right)$, on $h\to b\bar{b}$
\begin{equation}
BR\left(b\bar{b}\right)\equiv\frac{\Gamma\left(h_{SM}\to b\bar{b}\right)}{\Gamma\left(h_{SM}\to b\bar{b}\right)+\Gamma\left(h_{SM}\to H^{0}H^{0}\right)+\Gamma\left(h_{SM}\to A^{0}A^{0}\right)}<BR_{max}\left(b\bar{b}\right).\end{equation}

\item $Z$-decay constraint \begin{equation}
m_{Z}<m_{H^{0}}+m_{A^{0}}.\end{equation}

\item Vacuum stability bound given in Eq.\ \ref{eq:vacuumbound}. 
\end{itemize}
We note that the vacuum stability bound is highly correlated with
the bound on the decay of $h\to b\bar{b}$. It can be understood as
follows. Due to the $Z$-pole constraint $m_{H^{0}}$ should be larger
than $41\,{\rm GeV}$ for $\Delta m_{A^{0}H^{0}}=10\,{\rm GeV}$,
therefore, the SM-like Higgs boson has to be heavier than $82\,{\rm GeV}$
in order to open the $h\to H^{0}H^{0}$ mode. Bearing in mind the
fact that $BR_{max}\left(b\bar{b}\right)<0.1$ for a light SM-like
Higgs boson, e.g. $m_{h}\sim100\,{\rm GeV}$, the LEP II bound on
$h\to b\bar{b}$ now becomes \begin{equation}
\frac{3m_{b}^{2}m_{h}}{8\pi v^{2}}\beta_{b}^{3/2}<BR_{max}\left(b\bar{b}\right)\frac{v^{2}}{16\pi m_{h}}\left(\frac{m_{H^{0}}-\mu_{2}^{2}}{v^{2}}\right)^{2}\beta_{H^{0}},\end{equation}
 where $\beta_{b}=\sqrt{1-4m_{b}^{2}/m_{h}^{2}}$ and  $\beta_{H^{0}}=\sqrt{1-4m_{H^{0}}^{2}/m_{h}^{2}}$.
As $\beta_{H}\sim0$ for a light Higgs boson, $\mu_{2}$ has to be
very large, even larger than $m_{H^{0}}$, in order to satisfy the
above inequality. On the other hand, $\mu_{2}$ is also bounded from
above due to vacuum stability. It thus leads to an interesting connection
between the vacuum stability bound and the LEP II bound on $h\to b\bar{b}$.

\begin{figure}
\includegraphics[scale=0.5]{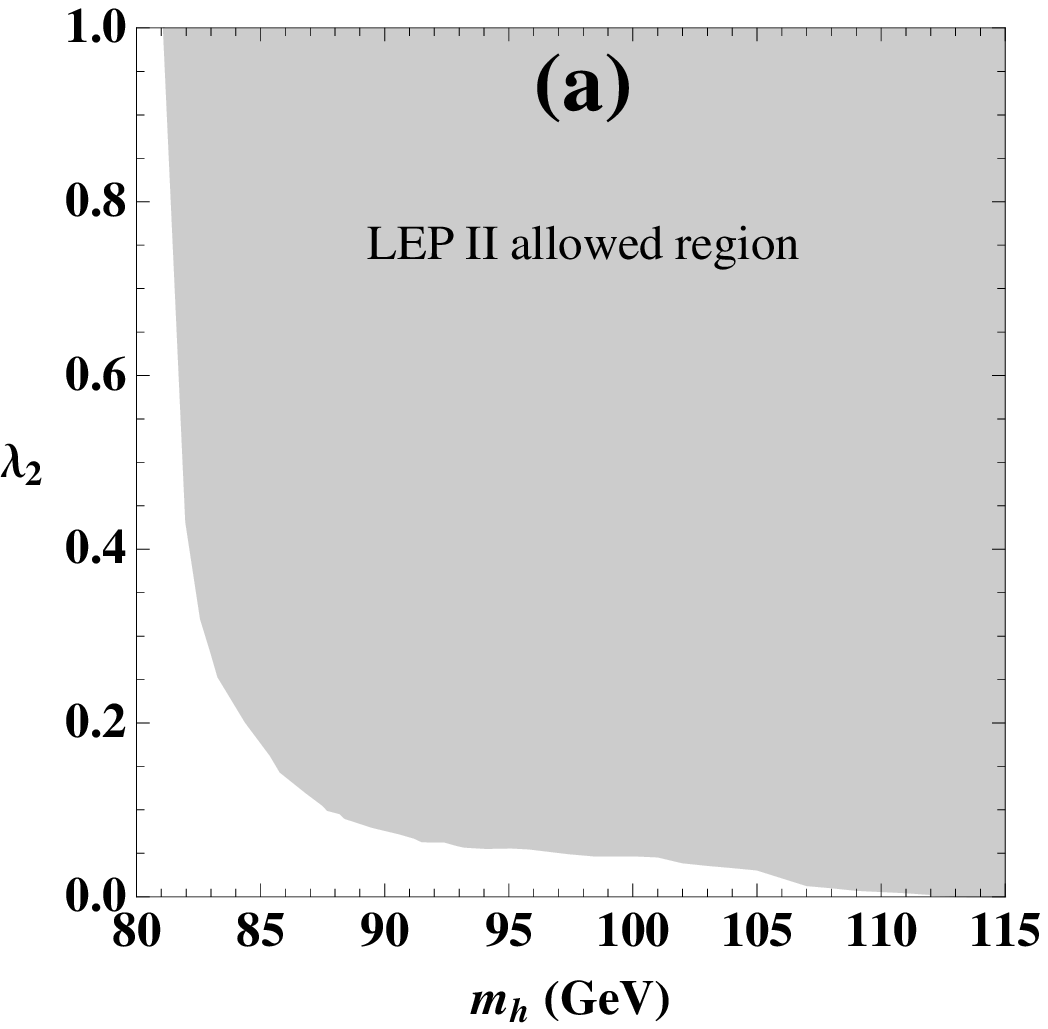}\ \ \includegraphics[scale=0.5]{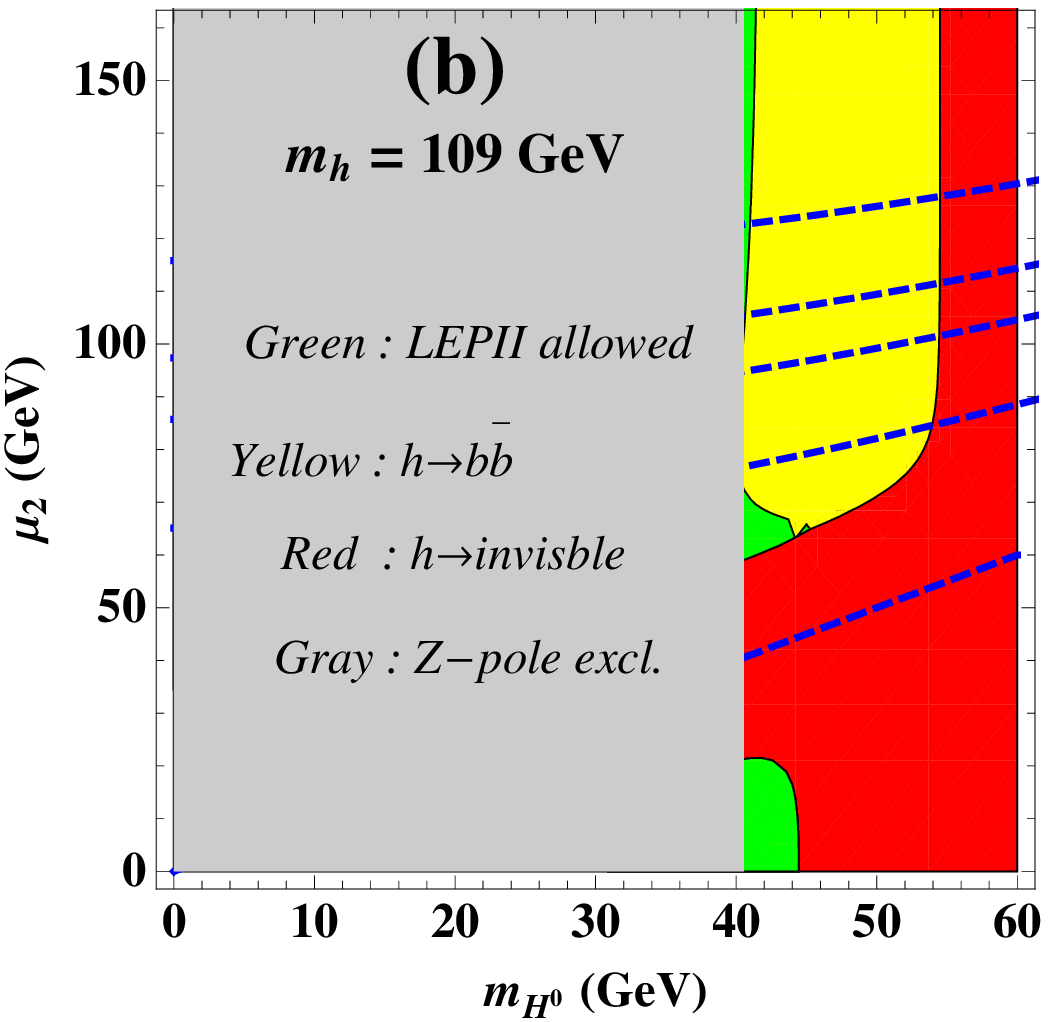}\ \ \includegraphics[scale=0.5]{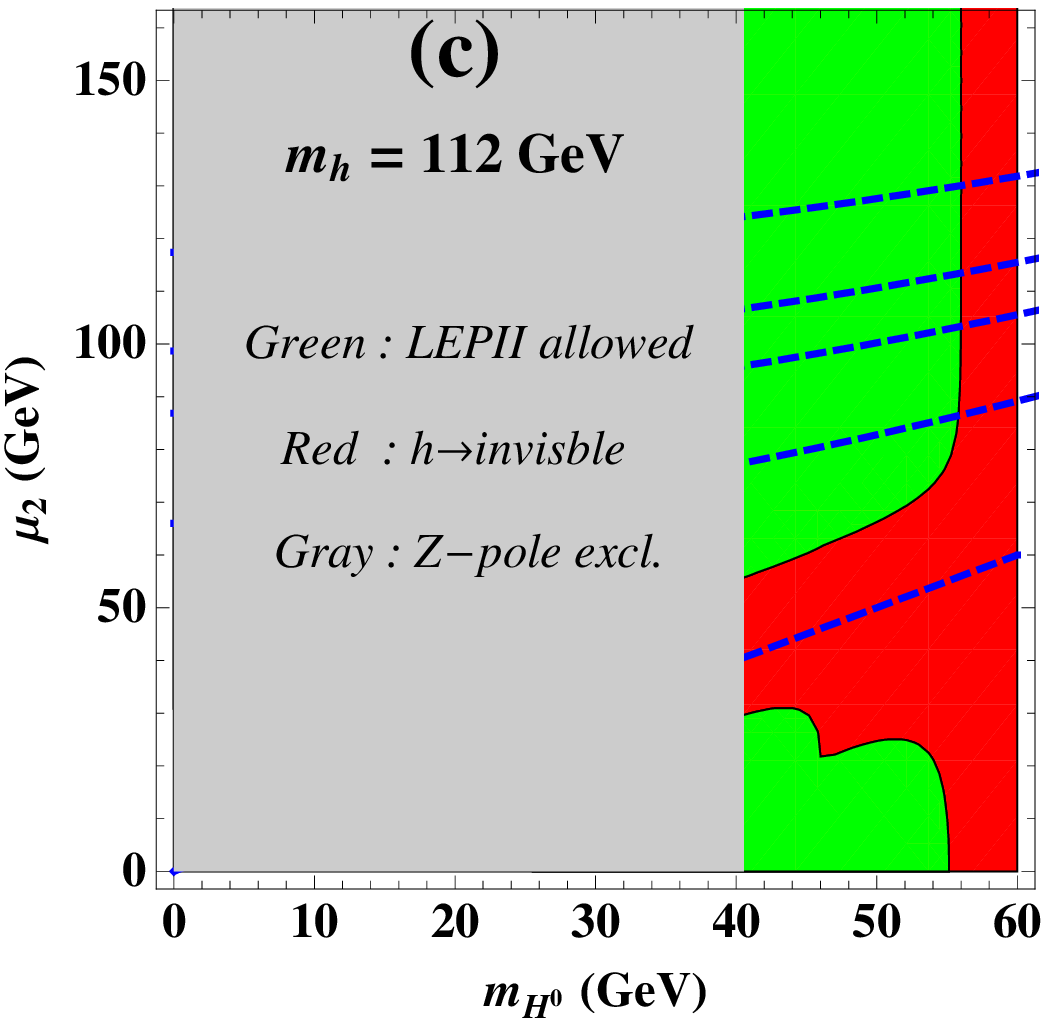}

\caption{(a) Allowed parameter space (gray region) of $(m_{h},\lambda_{2})$;
(b)(c) allowed parameter space of $\left(m_{H^{0}},\mu_{2}\right)$
after imposing the LEP II constraints on the decay branching ratio
of $h\to{\rm invisible}$ and $h\to b\bar{b}$. The region above the
dashed line is excluded by the vacuum stability requirement. From
top to bottom: $\lambda_{2}=1.0,\,0.5,0.3,\,0.1$and $\lambda_{2}\to0$.
\label{fig:allowed-parameter}}

\end{figure}

The allowed region of $\lambda_{2}$, which satisfies the constraints
on both $BR_{max}\left(inv\right)$ and $BR_{max}\left(b\bar{b}\right)$,
is shown in Fig.\ \ref{fig:allowed-parameter}(a). For the sake of
illustration, we plot the allowed parameter space of $\left(m_{H^{0}},\,\mu_{2}\right)$
for $m_{h}=109\,{\rm GeV}$ and $112\,{\rm GeV}$ in Fig.\ \ref{fig:allowed-parameter}(b)\ and\ (c).
In the DSDM the lower limits of $m_{h}$ are shown in Table\ \ref{tab:mhiggs_lep}
for various $\Delta m_{A^{0}H^{0}}$.

\begin{table}
\caption{Lower limits on the mass of the SM Higgs boson derived from LEP II
Higgs boson search for various $\Delta m_{A^{0}H^{0}}$. \label{tab:mhiggs_lep}}

\begin{tabular}{ccccccccccc}
\hline 
$\Delta m_{A^{0}H^{0}}\,\left({\rm GeV}\right)$ & \ \ \ \  & $5$ & \ \ \ \  & 10 & \ \ \ \  & 15 & \ \ \ \  & 20  & \ \ \ \  & 30\tabularnewline
\hline 
$m_{h}\,\left({\rm GeV}\right)$ &  & 105 &  & 106 &  & 109 &  & 110 &  & 110\tabularnewline
\hline
\end{tabular}
\end{table}

\section{Impact on the SM Higgs search at colliders}

\begin{figure}
\includegraphics[clip,scale=0.55]{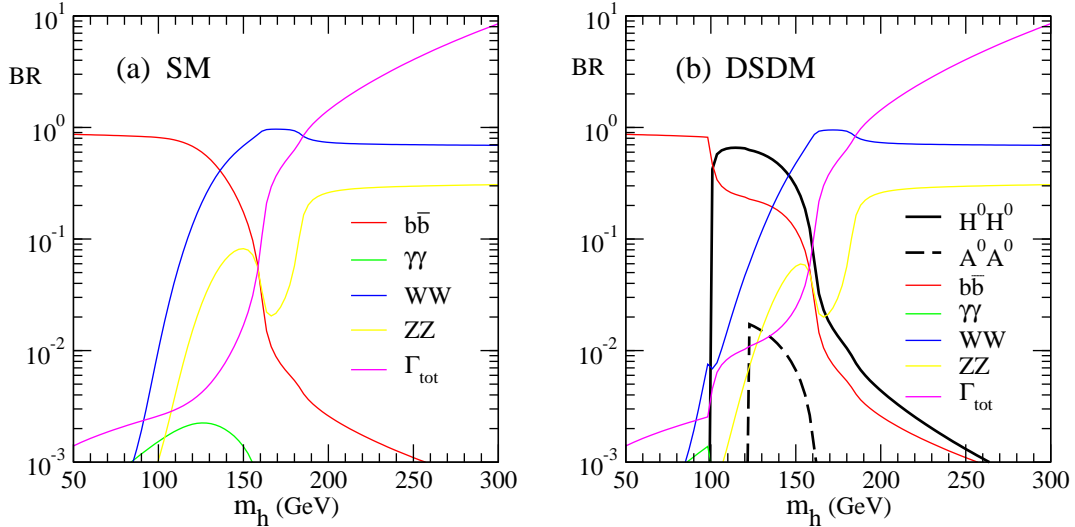}

\caption{(a) Selected Higgs boson decay branching ratios as a function of $m_{h}$
in the SM; (b) selected SM Higgs boson decaying ratios as a function
of $m_{h}$ in the DSDM. Here we have chosen $m_{H^{0}}=50\,{\rm GeV}$,
$\Delta m_{A^{0}H^{0}}=10\,{\rm GeV}$, $m_{H^{\pm}}=170\,{\rm GeV}$
and $\mu_{2}=20\,{\rm GeV}$. The vertical axis is units of GeV for
the total decay width. \label{fig:decaywidth}}

\end{figure}

At the LHC, the SM Higgs boson is mainly produced through gluon-gluon
fusion induced by a heavy (top) quark loop. Once produced, it will
decay into a fermion pair or vector boson pair. The strategy of searching
for the Higgs boson depends on how it decays and how large the decay
branching ratio is. We present selected decay branching ratios of
the SM Higgs boson as a function of $m_{h}$ in Fig.\ \ref{fig:decaywidth}(a),
as well as the total decay width ($\Gamma_{tot}$) in units of GeV.
If the SM Higgs boson is lighter than $130\,$GeV, it decays mainly
into a bottom quark pair ($b\bar{b}$). Unfortunately, it is very
difficult to search for the Higgs boson in this mode due to the extremely
large Quantum Chromodynamics (QCD) background at the LHC. However,
the $h\to\gamma\gamma$ mode can be used to detect a Higgs boson with
the mass below 150~GeV~\citep{Froidevaux:1995,Gianotti:1996} even
though the decay branching ratio of this mode is quite small, $\sim O(10^{-3})$.
If the Higgs boson mass ($m_{h}$) is in the region of $130\,{\rm GeV}$
to $2m_{Z}$, the $h\to ZZ^{*}$ mode is very useful because of its
clean collider signature of four isolated charged leptons. The $h\to WW^{(*)}$
mode is also important in this mass region because of its large decay
branching ratio\ \citep{Kauer:2000hi}. if $m_{h}>2m_{Z}$, the decay
mode $h\to ZZ\to\ell^{+}\ell^{-}\ell^{\prime+}\ell^{\prime-}$ is
considered as the {}``gold-plated'' mode which is the most reliable
way to detect the Higgs boson up to $m_{h}\sim600\,$GeV because the
backgrounds are known rather precisely and the two on-shell $Z$ bosons
could be reconstructed experimentally. For $m_{h}>600\,{\rm GeV}$,
one can detect the $h\to ZZ\to\ell^{+}\ell^{-}\nu\bar{\nu}$ decay
channel in which the signal appears as a Jacobian peak in the missing
transverse energy spectrum\ \citep{Cao:2007du}. In this section
we show a few representive cases to illustrate how the new dark scalars
affect the Higgs search at the LHC. Our choice of masses for all the
particles is not necessarily constrained by precision electroweak
data \citep{Barbieri:2006dq,Gerard:2007kn}. The numerical calculation
is carried out with the help of the HDECAY program\ \citep{Djouadi:1997yw},
after the implementation of the new decay channels in the DSDM.

The new decay channels of the SM Higgs boson will increase its decay
width but decrease the decay branching ratios of the usual decay modes,
such as $b\bar{b}$, $WW$, $ZZ$ and $\gamma\gamma$, etc. In Fig.\ \ref{fig:decaywidth}(b)
we present the relevant decay branching ratios of the SM Higgs boson
as a function of its mass. Clearly, the usual decay modes are highly
suppressed once the $H^{0}H^{0}$ and $A^{0}A^{0}$ decay modes open.
The invisible mode ($h\to H^{0}H^{0}$) dominates in the light and
intermediate mass region of $m_{h}$, i.e. $100<m_{h}<160\,{\rm GeV}$.
Due to the $10\,{\rm GeV}$ mass gap between $H^{0}$ and $A^{0}$,
the contribution of the $H^{0}H^{0}$ mode is much larger than that
of the $A^{0}A^{0}$ mode, as shown by the black solid and dashed
curves in Fig.\ \ref{fig:decaywidth}(b). It implies that the SM
Higgs boson decay depends mainly on $m_{H^{0}}$ but not $m_{A^{0}}$.
We also note that the $\gamma\gamma$ mode is suppressed so much that
it will be very challenging to use this mode to detect the SM Higgs
boson in the mass region of $m_{h}\sim100-150\,{\rm GeV}$.

\begin{figure}
\includegraphics[clip,scale=0.65]{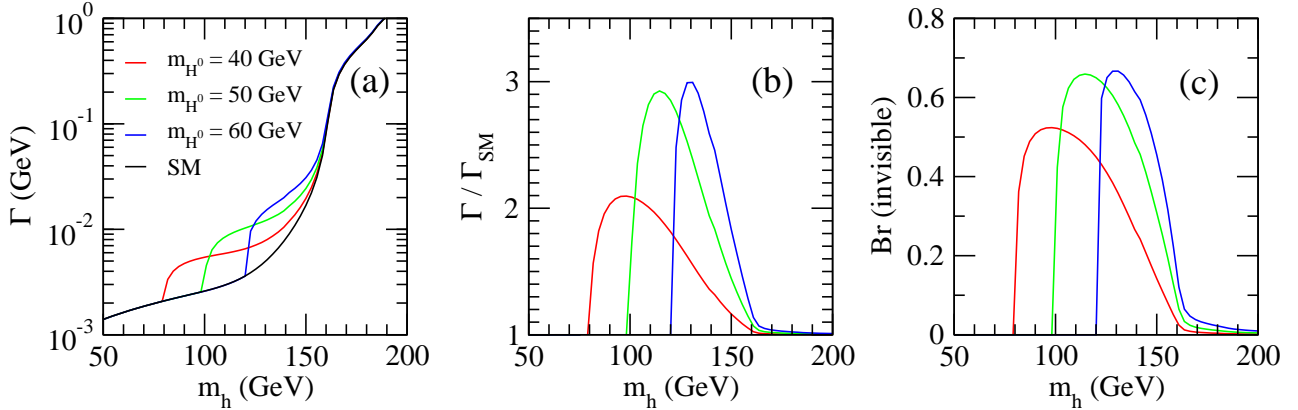}

\caption{(a) Total decay width of the SM Higgs boson as a function of $m_{h}$;
(b) ratio of the total decay width of the SM Higgs boson in the DSDM
and Higgs boson in the SM; (c) decay branching ratio of the invisible
decay mode of the SM Higgs boson in the DSDM. For comparison, we choose
$m_{H^{0}}=40\,\left(50,\,60\right)\,{\rm GeV}$, $\Delta m_{A^{0}H^{0}}=10\,{\rm GeV}$,
$m_{H^{\pm}}=170\,{\rm GeV}$, and $\mu_{2}=20\,{\rm GeV}$. \label{fig:decaywidth2}}

\end{figure}

To examine the dependence of the SM Higgs decay upon $m_{H^{0}}$,
we choose sample points of $m_{H^{0}}$ being $40\,\left(50,\,60\right)\,{\rm GeV}$
and present the corresponding Higgs boson total decay width, the ratio
of total decay width in the DSDM to the one in the SM ($\Gamma/\Gamma_{SM}$)
and the decay branching ratio of the invisible decay mode ($Br\left(invisible\right)$)
in Fig.\ \ref{fig:decaywidth2}(a)\ (b)\ and\ (c), respectively.
Clearly, the total decay width increases by a big amount once the
new decay channel opens. For example, as shown in Fig.\ \ref{fig:decaywidth2}(b),
the new invisible decay mode can enlarge the total width by a factor
of three for $m_{H^{0}}=60\,{\rm GeV}$ and $m_{h}=130\,{\rm GeV}$
and by a factor of two for $m_{H^{0}}=40\,{\rm GeV}$ and $m_{h}=100\,{\rm GeV}$.
The decay branching ratio of the invisible mode is about $50-65\%$
in the mass region $m_{h}\sim100-150\,{\rm GeV}$ for $m_{H^{0}}\sim40-60\,{\rm GeV}$.
Needless to say, the $WW$ mode dominates over the invisible mode
once the SM Higgs boson is heavier than $160\,{\rm GeV}$, beyond
which the SM Higgs boson in the DSDM is essentially the same as that
of the SM.

\begin{figure}
\includegraphics[clip,scale=0.6]{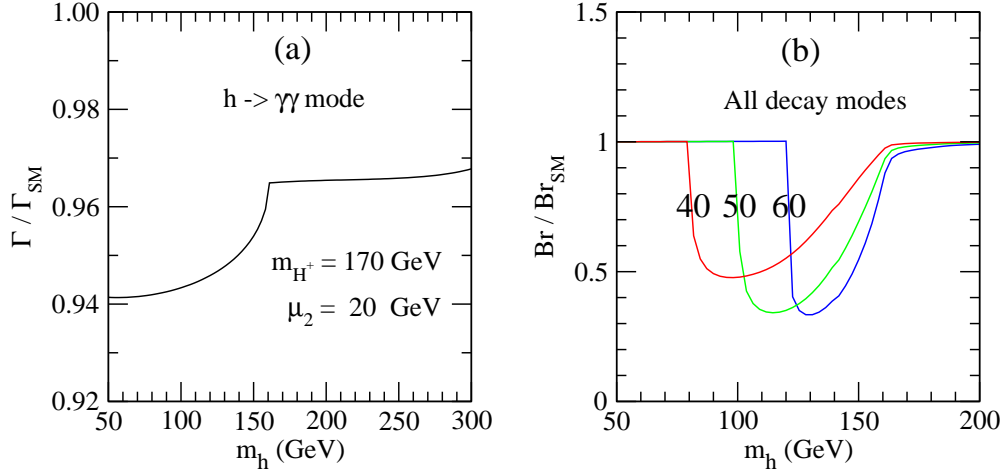}

\caption{(a) Ratio of the decay branching ratios of $h_{SM}\to\gamma\gamma$
in the DSDM and the SM; (b) suppression factor of the usual decay
modes compared to the SM for $m_{H^{0}}=40,\,50,\,60\,{\rm GeV}$.
Here we have chosen $\Delta m_{A^{0}H^{0}}=10\,{\rm GeV}$ and $\mu_{2}=20\,{\rm GeV}$.
\label{fig:ratio-decbr}}

\end{figure}

\newpage{} In the DSDM the decay mode of $h\to\gamma\gamma$ receives
an additional contribution from the charged Higgs boson loop. The
partial decay width of $h\to\gamma\gamma$ is given by\ \citep{Djouadi:2005gi,Djouadi:2005gj}
\begin{eqnarray*}
 &  & \Gamma\left(h\to\gamma\gamma\right)\\
 & = & \frac{G_{\mu}\alpha^{2}m_{h}^{3}}{128\sqrt{2}\pi^{3}}\left|\sum_{f}N_{C}Q_{f}^{2}g_{hff}\mathcal{A}_{1/2}\left(\tau_{f}\right)+g_{hWW}\mathcal{A}_{1}\left(\tau_{W}\right)+\frac{m_{H^{\pm}}^{2}-\mu_{2}^{2}}{\sqrt{2}\, m_{H^{\pm}}^{2}}\,\mathcal{A}_{0}\left(\tau_{H^{\pm}}\right)\right|^{2},\end{eqnarray*}
 where the amplitude $\mathcal{A}_{i}$ and coupling $g_{hff/hWW}$
are given in Refs.\ \citep{Djouadi:2005gi,Djouadi:2005gj}. For the
numerical calculation we also implemented the correction from the
charged Higgs boson loop in the HDECAY program. The impact of the
invisible decay on the other regular decay modes are shown in Fig.\ \ref{fig:ratio-decbr}.
We note that the loop correction of the charged Higgs boson slightly
suppresses the partial decay width of $h\to\gamma\gamma$ by about
$10\%$, as shown in Fig.\ \ref{fig:ratio-decbr}(a). Such a suppression
makes it more difficult to search for the SM Higgs boson via the di-photon
mode. Furthermore, as an overall effect, all the usual decay modes
of the SM Higgs boson, e.g. $b\bar{b}/WW/ZZ/\gamma\gamma$, are highly
suppressed once the new decay mode opens. In the intermediate mass
regime, $130<m_{h}<160\,{\rm GeV}$, the suppression factor can be
larger than 0.5 which definitely makes the search of the SM Higgs
boson more challenging.

\begin{figure}
\includegraphics[scale=0.6]{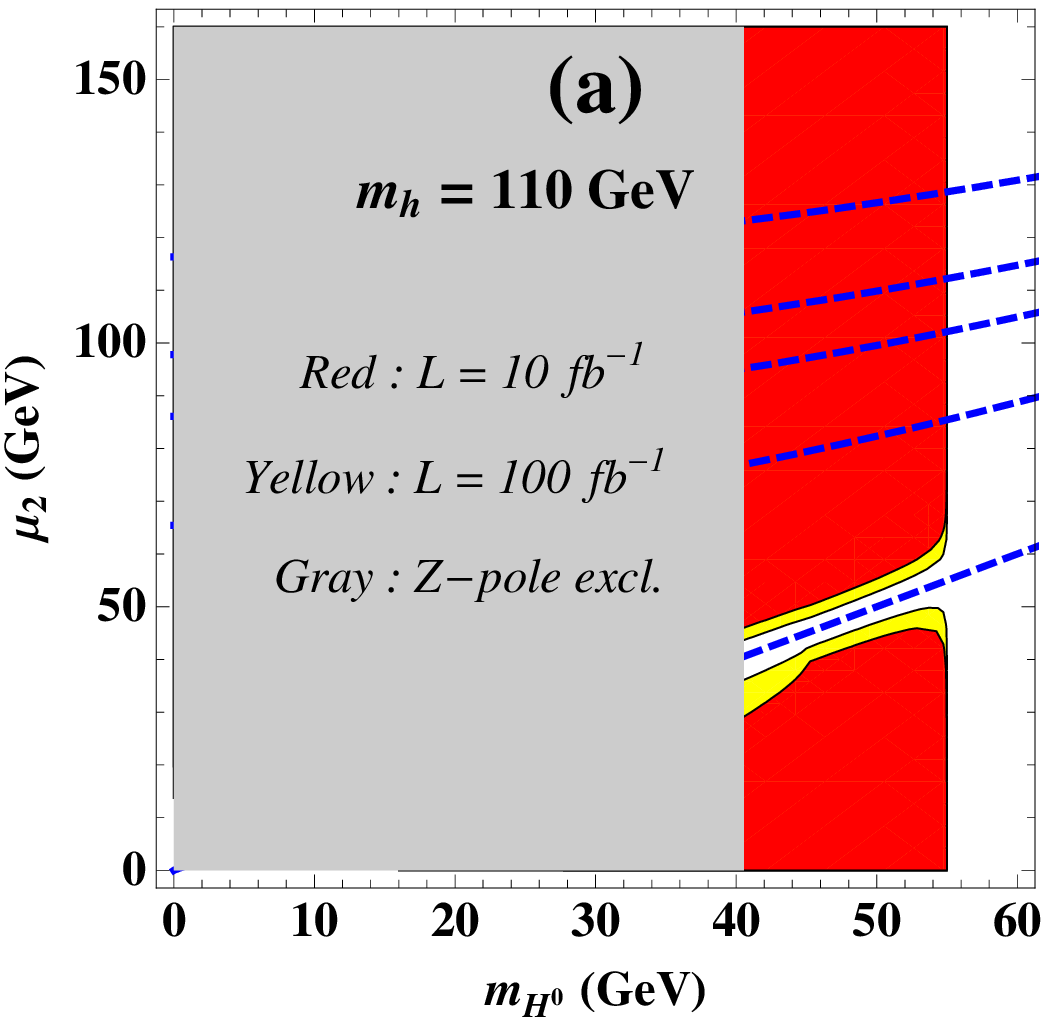}\ \ \ \ \ \ \includegraphics[scale=0.6]{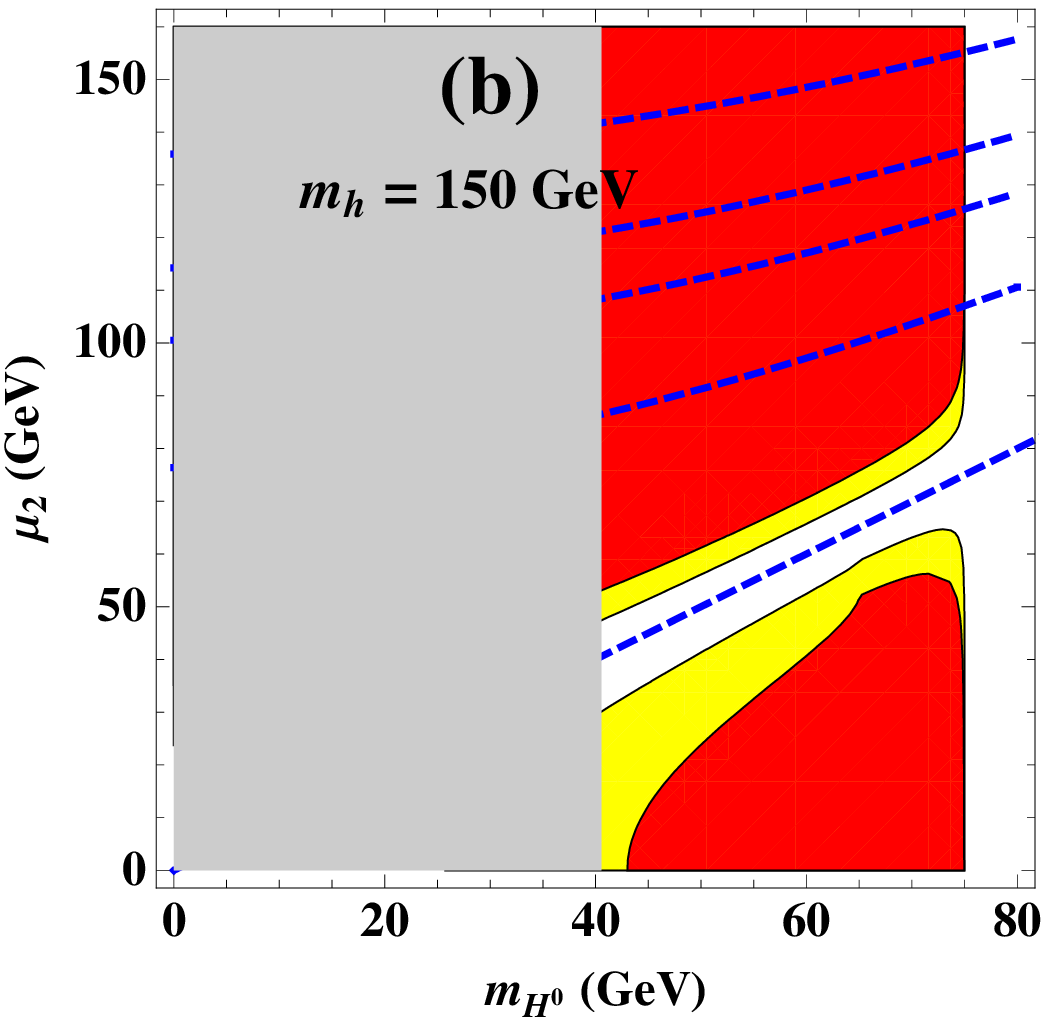}

\caption{ 
Allowed parameter space of $\left(m_{H^{0}},\mu_{2}\right)$ to reach
the $5\sigma$ discovery of the invisible Higgs decay via the WBF
process: (a) $m_{h}=110\,{\rm GeV}$, (b) $m_{h}=150\,{\rm GeV}$.
The region above the dashed-lines
is excluded by the vacuum stability requirement. From top to bottom:
$\lambda_{2}=1.0,\,0.5,\,0.3,\,0.1$ and $\lambda_{2}\to0$. 
\label{fig:WBF-seach}}

\end{figure}

On the other hand, one might search for the Higgs boson through its
invisible decay mode. For example, the weak-boson fusion (WBF) process,
$q\bar{q}\to q^{\prime}\bar{q^{\prime}}VV\to q^{\prime}\bar{q^{\prime}}h$
with the subsequential decay to undetectable particles, i.e. $h\to H^{0}H^{0}/A^{0}A^{0}$.
The signal is thus characterized by two quark jets, which typically
stay in the foward and backward regions of the detector and are widely
separated in pseudorapidity, and also by a large missing transverse
momentum ($\met$), due to the Higgs invisible decay products. It
has been shown in Ref.\ \citep{Eboli:2000ze} that the WBF process
can be used to detect the invisible decay Higgs boson at $5\sigma$
level, when statistical errors only are considered, if the invisible
decay branching ratio satisfies the following condition \begin{equation}
\frac{\sigma_{S}\times Br\left(h\to invisible\right)\times\mathcal{L}}{\sqrt{\sigma_{B}\times\mathcal{L}}}>5,\label{eq:wBF-search}\end{equation}
 where $\mathcal{L}$ denotes the integrated luminosity and $\sigma_{S}(\sigma_{B})$
the cross section of the signal (background) respectively. After imposing
the kinematical cuts suggested in Ref.\ \citep{Eboli:2000ze}, $\sigma_{B}=167\,{\rm fb}$
and $\sigma_{S}=99.4\,(99.7,\,94.3,\,89.2)\,{\rm fb}$ for $m_{h}=110\,(120,\,130,\,150)\,{\rm GeV}$.
From the above relation, we obtain the allowed parameter region of
$\left(m_{H^{0}},\mu_{2}\right)$ to reach the $5\sigma$ discovery
as shown in Fig.\ \ref{fig:WBF-seach}. Most of the parameter space
can be covered for $\mathcal{L}=100\,{\rm fb}^{-1}$, c.f. the yellow
region.

Unfortunately, finding or constraining an invisibly decaying Higgs
boson signal in $jj+\met$ event is essentially a counting experiment
since a resonance in the invariant mass distribution of the Higgs
decay products cannot be extracted. It also depends on two or more
independent model parameters, thus one can only probe or constrain
their combination. Next we are going to examine the $dark$ scalar
pair production at the LHC, which is complementary to the WBF process.

\section{Collider Phenomenology of the Dark Scalar Doublet}

\subsection{Production of dark scalar pair at the LHC}

\begin{figure}
\includegraphics[clip,scale=0.7,angle=270]{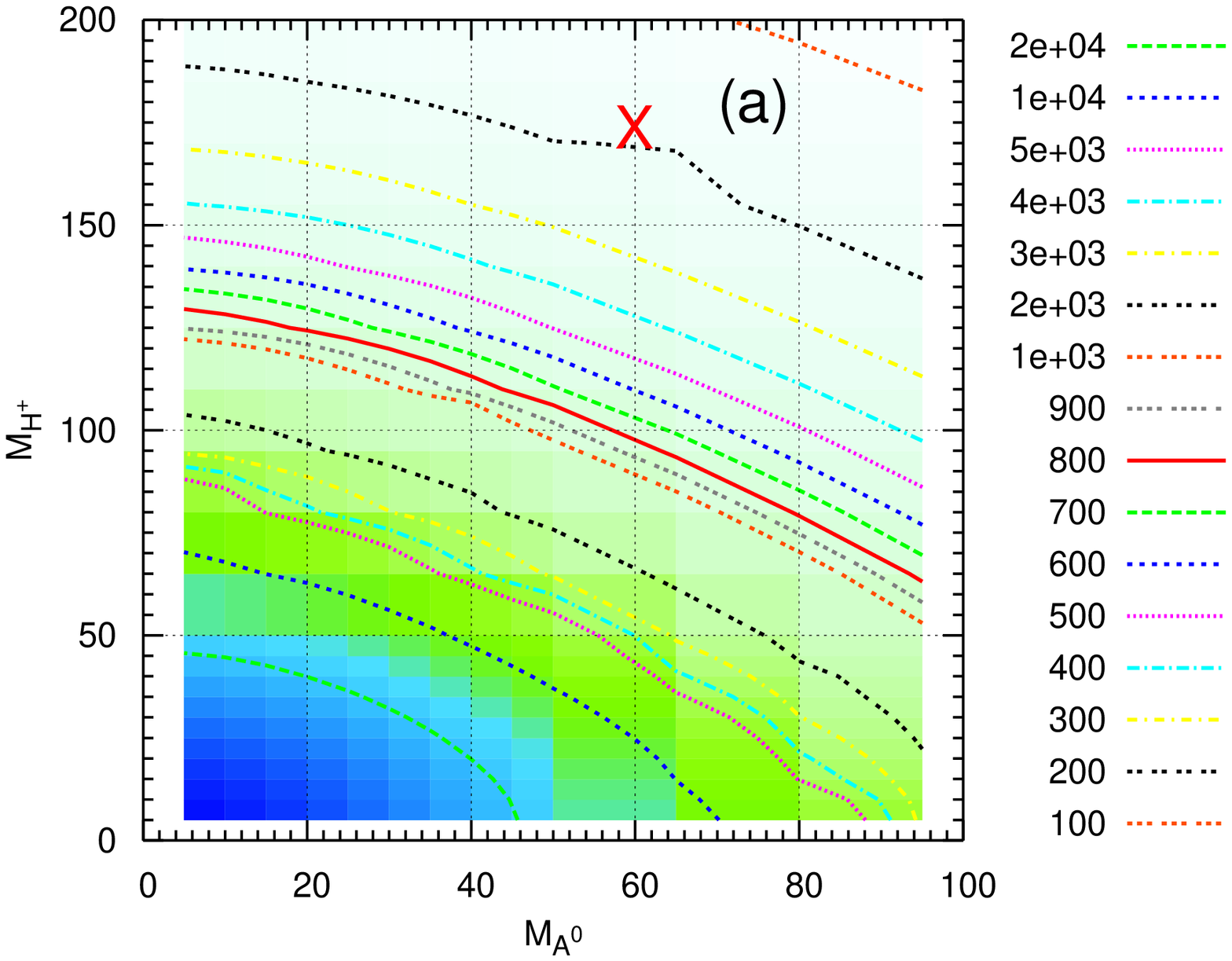}

\includegraphics[clip,scale=0.7,angle=270]{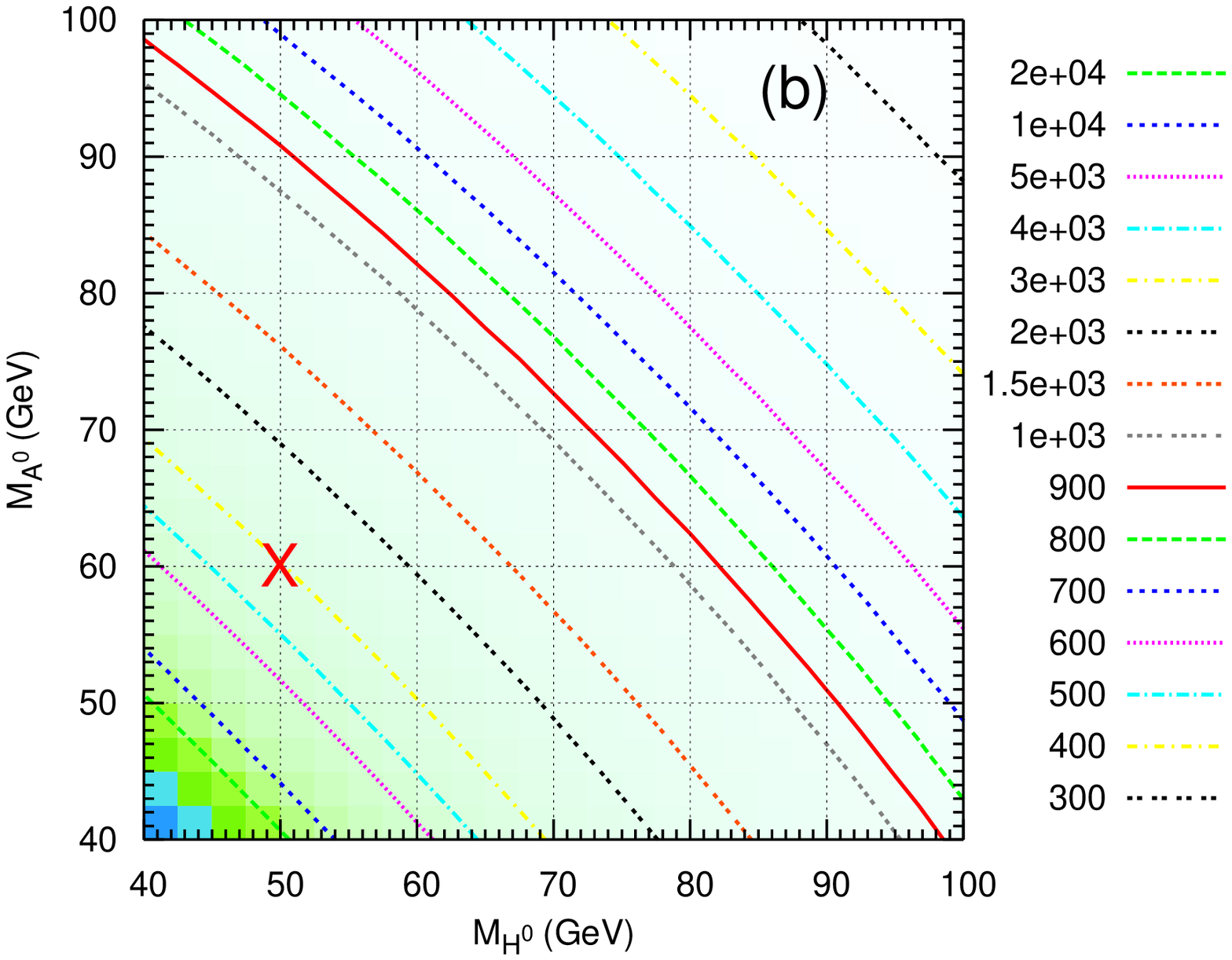}

\caption{Contours of the cross sections (in $fb$ units) of $A^{0}H^{+}$ production
(a) and of $A^{0}H^{0}$ production (b) at the LHC in the plane of
the scalars' masses. The red symbol {}``X'' denotes our benchmark
point used in the collider study. Note that Fig.\ (a) also holds
for $H^{0}H^{\pm}$ production. \label{fig:xsec-wah}}

\end{figure}

Since all new scalars carry a $Z_{2}$-odd quantum number, they cannot
be produced singly at the collider. Furthermore, these new scalars
only couple to the Higgs boson and electroweak gauge bosons of the
Standard Model. Hence they should be produced via processes of the
Drell-Yan type at the LHC as listed below: \begin{eqnarray}
 &  & q\bar{q^{\prime}}\to W^{*}\to A^{0}H^{\pm},\label{eq:A0HC}\\
 &  & q\bar{q^{\prime}}\to W^{*}\to H^{0}H^{\pm},\label{eq:H0HC}\\
 &  & q\bar{q}\to Z^{*}\to A^{0}H^{0},\label{eq:A0H0}\\
 &  & q\bar{q}\to Z(\gamma)\to H^{+}H^{-}.\label{eq:HCHC}\end{eqnarray}
 In the DSDM, the vertex of $V-\phi-\phi^{'}$ is a pure gauge coupling,
where $V$ denotes an SM gauge boson while $\phi(\phi')$ is a $Z_{2}$-odd
scalar. Therefore, the total cross section of each production process
depends only on the masses of the $Z_{2}$-odd scalars in the final
state. In this work the matrix elements of both signal and background
processes are calculated using MADGRAPH\ \citep{Stelzer:1994ta,Maltoni:2002qb}
while the widths of the dark scalars are calculated in CALCHEP\ \citep{Pukhov:2004ca}
with the model file generated by LANHEP\ \citep{Semenov:2002jw}.
Agreement of both programs has been checked at the level of dark scalar
pair production. In Fig.\ \ref{fig:xsec-wah} we present the contours
of the cross sections (in ${\rm fb}$ units) of the $A^{0}H^{+}$
and $A^{0}H^{0}$ associated production at the LHC in the plane of
the scalars' masses. The red symbol {}``X'' denotes one of our benchmark
pionts, $\left[m_{H^{0}},\, m_{A^{0}},\, m_{H^{\pm}}\right]=\left[50,\,60,\,175\right]\,{\rm GeV}$,
used in the collider study. We note that the production cross section
of the dark scalar pair at the LHC is sizable, e.g. at the benckmark
point shown in Fig.\ \ref{fig:xsec-wah} the cross section of the
$A^{0}H^{0}$ pair production is about $3\,{\rm pb}$ while that of
the $A^{0}H^{\pm}$ pair production is about $200\,{\rm fb}$. The
latter is highly suppressed owing to the large mass of $H^{\pm}$.
Due to the same reason the cross section of the $H^{+}H^{-}$ pair
production is quite small in the relevant parameter space (around
the benchmark piont); it is therefore not shown in this study.

\subsection{Decay pattern of $A^{0}$ and $H^{\pm}$}

\begin{figure}
\includegraphics[clip,scale=0.4]{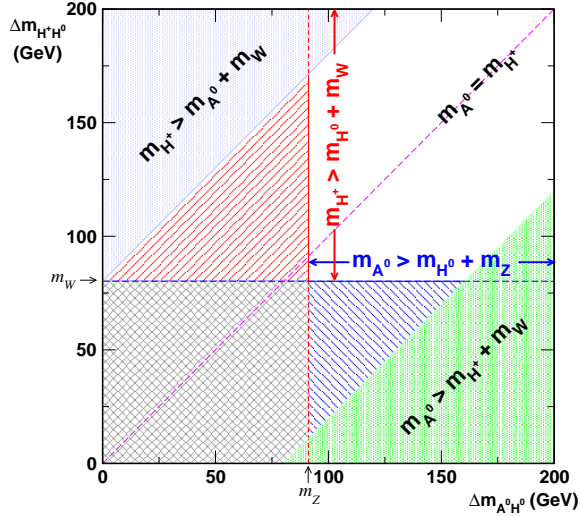}

\caption{Mass pattern that determines the decay pattern of $A^{0}$ and $H^{+}$.\label{fig:Decay-pattern}}

\end{figure}

The decay pattern of $A^{0}$ and $H^{\pm}$ is completely determined
by three mass spacings $\Delta m_{A^{0}H^{0}}$, $\Delta m_{H^{+}H^{0}}$
and $\Delta m_{H^{+}A^{0}}$, as shown in Fig.\ \ref{fig:Decay-pattern},
where $\Delta m_{H^{+}H^{0}}\equiv m_{H^{+}}-m_{H^{0}}$ and $\Delta m_{H^{+}A^{0}}\equiv m_{H^{+}}-m_{A^{0}}$.
Using the constraint from precision electroweak data, it has been
shown by Ref.\ \citep{Barbieri:2006dq} that a relationship exists
between the first and second spacing in the $Z_{2}$-odd scalar spectrum\begin{equation}
\left(m_{H^{+}}-m_{A^{0}}\right)\left(m_{H^{+}}-m_{H^{0}}\right)=M^{2},\qquad M=120_{-30}^{+20}\,{\rm GeV},\label{eq:ewpt}\end{equation}
 which implies that the charged Higgs boson should be heavier than
both $A^{0}$ and $H^{0}$ so that $A^{0}$ can only decay into the
$H^{0}Z^{*}$ mode. This is consistent with our choice. It corresponds
to the top-left corner in Fig.\ \ref{fig:Decay-pattern}. Of course,
the precision electroweak constraint also requires the SM Higgs to
be very heavy in this case, say $m_{h}>400$ GeV, but that does not
affect the production and decay of the dark scalar doublet particles
which are independent of $m_{h}$. {[}Our view is that the masses
of all the scalars being considered here are not necessarily constrained
by precision electroweak data or dark matter relic abundance because
there may be other contributions. We simply explore a set of reasonable
possibilities.]

\begin{figure}
\includegraphics[clip,scale=0.5]{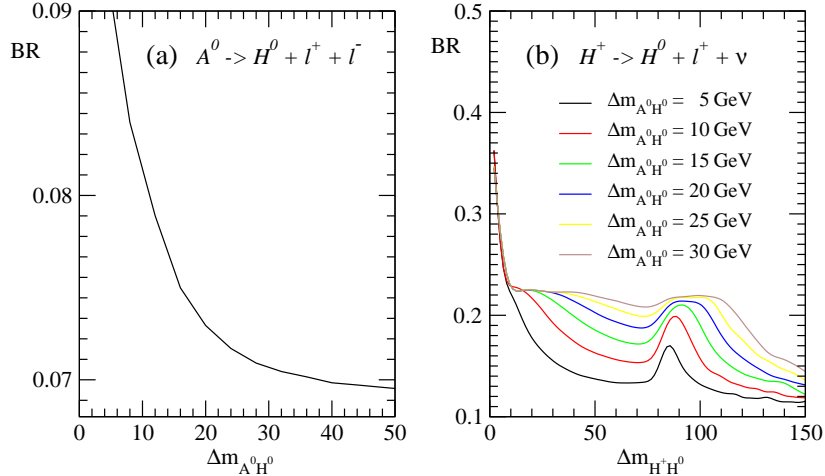}

\caption{Decay branching ratio of the $A^{0}\to H^{0}\ell^{+}\ell^{-}$ and
the $H^{+}\to H^{0}\ell^{+}\nu$ where we have summed over two lepton
flavors (electron and muon). Note that $\Delta m_{H^{+}H^{0}}$ and
$\Delta m_{A^{0}H^{0}}$ should satisfy the electroweak precision
test bound given in Eq.\ \ref{eq:ewpt}.\label{fig:decayBR}}

\end{figure}

Furthermore, in order to avoid the huge QCD (quantum chromodynamics)
background at any hadron collider, one has to use the leptonic decay
mode to tag the signal event. Thus, in Fig.\ \ref{fig:decayBR} we
plot the decay branching ratio (BR) of $A^{0}\to H^{0}\ell^{+}\ell^{-}$
and $H^{+}\to H^{0}\ell^{+}\nu$ as a function of $\Delta m_{A^{0}H^{0}}$
and $\Delta m_{H^{+}H^{0}}$ respectively. In these figures we have
summed over two lepton flavors (electron and muon). We note that $BR\left(A^{0}\to H^{0}\ell^{+}\ell^{-}\right)$
is only sensitive to $\Delta m_{A^{0}H^{0}}$. If the mass spacing
is small, say $\Delta m_{A^{0}H^{0}}\sim10\,{\rm GeV}$, the BR is
about 0.09. As the mass spacing increases, the BR decreases because
the massive quark modes, e.g. $s\bar{s}\,(c\bar{c},b\bar{b})$, increase.
For a very large mass spacing, the BR approaches to the BR of an on-shell
$Z$-boson. The decay of $H^{+}$ is a little complicated because
it can decay into two modes, $H^{0}W^{(*)}$ or $A^{0}W^{(*)}$. For
simplicity, we only focus on the decay mode $H^{+}\to H^{0}W^{(*)}$
in this study because the $H^{+}\to A^{0}W^{(*)}$ mode involves several
particles in the final state which render the analysis of the collider
signals much more complicated. Again, we note that the BR of $H^{+}\to H^{0}\ell^{+}\nu$
decreases as the mass spacing $\Delta m_{H^{+}H^{0}}$ increases because
of the massive quark modes. The peaks in the figure are due to the
mass threshold of the $W$-boson. Eventually, the BR approaches the
SM $W$-boson decay branching ratio, $\sim1/9$, at the large mass
spacing regime.

In the following sections, we present the results of a detailed Monte
Carlo simulation and determine the discovery potential of the $Z_{2}$-odd
scalars at the LHC. We find that the $A^{0}H^{+}/H^{0}H^{+}/H^{+}H^{-}$
pair productions suffer from huge backgrounds such that the dark scalar
signals are always overwhelmed. This difficulty persists even if we
lower the mass of $H^{\pm}$. We thus focus on the $A^{0}H^{0}$ associated
production in the next section.

\subsection{$A^{0}H^{0}$ production and signature}

The neutral scalar bosons can be produced via the following process
\begin{equation}
pp\to Z\to A^{0}H^{0},~~A^{0}\to H^{0}Z^{(*)},~~Z^{(*)}\to ff,\label{eq:ppA0H0}\end{equation}
 which gives rise to a collider signature as $ff+\met$. Thus, the
dominant SM background comes from two processes: \begin{eqnarray}
 &  & pp\to ZZ/Z\gamma,~{\rm with}~Z\to\nu\bar{\nu},~~Z/\gamma\to\ell^{+}\ell^{-},\label{eq:ZZ}\\
 &  & pp\to W^{+}W^{-},~{\rm with}~W^{+}\to\ell^{+}\nu,~~W^{-}\to\ell^{-}\bar{\nu},\label{eq:WW}\end{eqnarray}
 where two charged leptons in the $WW$ background are required to
have the same lepton flavors. Besides these two intrinsic (irreducible)
sources, there are also other reducible sources of background such
as $t\bar{t}$ pair production and $Wt$ associated production. For
example, if the bottom quark from top quark decay escapes the detector,
they will contribute to the mssing transverse momentum and thus mimic
the signal events. But one can suppress these reducible backgrounds
by vetoing any additional jet activity in the central rapidiy region.
Hence, we focus our attention only on the intrinsic background hereafter.

Below we choose a benchmark point, $\left[m_{H^{0}},m_{A^{0}},m_{H^{+}}\right]=\left[50,\,60,\,170\right]\,{\rm GeV}$,
to illustrate how to discover such a signature at the LHC. In order
to mimic the detector, we first require both leptons in the final
state to satisfy the following \emph{basic cuts}: \begin{eqnarray}
p_{T}^{\ell} & \geq & 15\,{\rm GeV},\qquad\left|\eta^{\ell}\right|\leq3.0,\label{eq:basic-cut}\end{eqnarray}
 where $p_{T}^{\ell}$($\eta^{\ell}$) denotes the transverse momentum
(rapidity) of the charged lepton $\ell$. Here we have assumed a perfect
detector that can precisely measure the four-momenta of the final-state
leptons. The number of signal and background events, after imposing
the basic cut, is given in the second column of Table\ \ref{tab:Number-of-event}.
The number of signal events is much less than the one of background
events, e.g. the ratio of the signal and background event ($S/B$)
is about $10^{-3}$ where $S$ and $B$ denotes the number of the
signal and background events, respectively. Obviously, additional
kinematical cuts are needed to enhance the ratio of signal to background.
For that, we examine the kinematics differences between signal and
backgroud below and find out the optimal kinematical cuts to entangle
the signal out of the background.

\begin{figure}
\includegraphics[clip,scale=0.6]{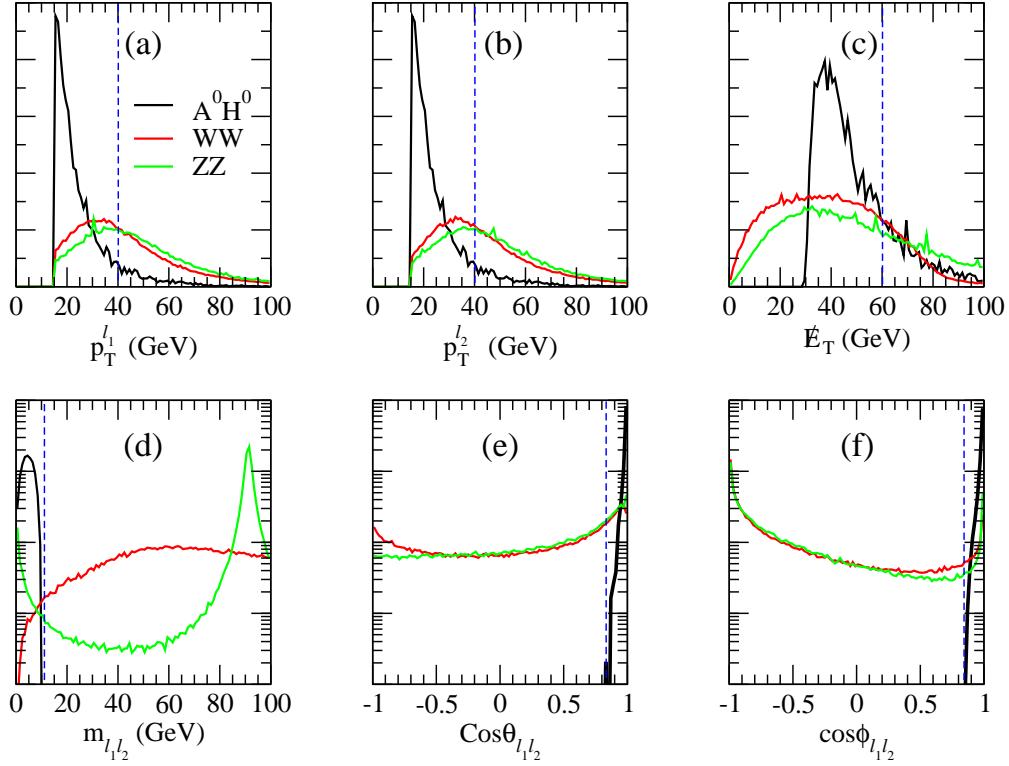}

\caption{(Normalized) kinematic distributions of the $A^{0}H^{0}$ production
for the benchmark point $\left[m_{H^{0}},m_{A^{0}},m_{H^{+}}\right]=\left[50,\,60,\,170\right]\,{\rm GeV}$
(black curve), the $WW$ process (red curve) and the $ZZ$ process
(green curve) after imposing the basic cuts. The blue dashed-line
indicates the optimal kinematical cuts which can be used to suppress
the SM background. \label{fig:A0H0-kinematics}}

\end{figure}

In Fig.\ \ref{fig:A0H0-kinematics} we present (normalized) kinematics
distributions of both the signal event and the background event after
imposing the basic cuts, where $\theta_{\ell\ell}$($\phi_{\ell\ell}$,
$m_{\ell\ell}$) denotes the open angle (azimuthal open angle, invariant
mass) of the two charged leptons and $\met$ the missing transverse
momentum originated from the two dark scalars in the final state.
We note that the kinematics of the signal events significantly differs
from the one of background events in the following aspects:

\begin{itemize}
\item Since the two charged leptons in the signal event come from an off-shell
$Z$-boson decay, they populate the small $p_{T}$ region and they
prefer to move parallel to each other, see Fig.\ \ref{fig:A0H0-kinematics}(a,b,e,f).
On the other hand, the charged leptons of the background events have
large $p_{T}$ and large open angles. 
\item Since the two missing scalars in the signal events are massive, one
may expect a large $\met$, but unfortunately, the two missing scalars
prefer to move back-to-back and their combination actually results
in a small $\met$ distribution, see Fig.\ \ref{fig:A0H0-kinematics}(c).
The distribution of the $ZZ$ background peaks around $40\,{\rm GeV}$
while the one of the $WW$ background peaks in the large $\met$ region. 
\item The invariant mass distribution of the two charged leptons of the
signal events peaks around $\sim\frac{1}{2}\Delta m_{A^{0}H^{0}}$
as they come from the decay of an off-shell $Z$-boson. The $ZZ$
background distribution exhibits two peaks: one is around $m_{Z}$
(from the on-shell $Z$-boson decay) and the other is around zero
(from the virtual photon decay). The intermediate region between the
two peaks is due to the interference effects between the $ZZ$ and
$Z\gamma$ processes. The distribution of the $WW$ background peaks
in the large invariant mass regime. 
\end{itemize}
Taking advantage of the kinematic differences between the signal and
the background, one can impose several optimal kinematic cuts, indicated
by the blue dashed-lines in Fig.\ \ref{fig:A0H0-kinematics}, to
extract the signal from the background.

Based on the topological difference between the signal and background
events, we impose the following \emph{optimal cuts} to futher suppress
the background, \begin{eqnarray}
p_{T}^{\ell}\leq40\,{\rm GeV}, & \qquad & \met\leq60\,{\rm GeV},\nonumber \\
\cos\theta_{\ell\ell}\ge0.9, & \qquad & \cos\phi_{\ell\ell}\ge0.9.\label{eq:optimal-cut}\end{eqnarray}
 At last, we require that the invariant mass of the two charged leptons
satisfies the following \emph{mass window} cut: \begin{equation}
0\leq m_{\ell\ell}\leq10\,{\rm GeV}.\label{eq:masswindow-cut}\end{equation}
 In Table\ \ref{tab:Number-of-event}, we give the number of events
at the LHC for an integrated luminosity of 100 $(fb)^{-1}$ of the
processes metioned above as well as those of the backgound. The kinematic
cuts listed in each column are applied sequentially. The two charged
leptons in the final state can be either electron or muon. We take
it into account by summing over the two lepton species. We also summed
over three species of neutrinos in the $ZZ$ background. In order
to generalize our study, we also examine the collider reach when $\Delta m_{A^{0}H^{0}}=20\,{\rm GeV}$
and $30\,{\rm GeV}$. As shown in the table, a staitistical significance
$S/\sqrt{B}\ge3$ can be achieved after imposing the optimal cuts
and it is further increased after imposing the mass window cut for
$\Delta m_{A^{0}H^{0}}=10\,{\rm GeV}$ and $20\,{\rm GeV}$, but slightly
decreased for $\Delta m_{A^{0}H^{0}}=30\,{\rm GeV}$ because the mass
window cut, $m_{\ell\ell}<10\,{\rm GeV}$, also cuts away $\sim32\%$
signal events. With an integrated luminosity $300\,{\rm fb}^{-1}$,
all three cases can reach $5\sigma$ significance even before the
mass window cut, therefore, LHC has a great potential to observe the
dark scalar.

\begin{table}
\caption{Number of signal and background events, for the benchmark points $\left[m_{H^{0}},m_{A^{0}},m_{H^{+}}\right]=\left[50,\,60(70\,,\,80),\,170\right]\,{\rm GeV}$,
at the LHC with an integrated luminosity $100\,{\rm fb}^{-1}$. The
kinematic cuts listed in each column are applied sequentially. Note
that we summed over two species of charged lepton, i.e. electron and
muon, for both signal and background and we also summed over three
species of neutrinos in the $ZZ$ background. \label{tab:Number-of-event}}

\begin{tabular}{ccccccc}
\hline 
BKGD &  & basic &  & optimal &  & $m_{\ell\ell}<10\,{\rm GeV}$\tabularnewline
\hline 
$WW$ & $\qquad$ & $1.1\times10^{5}$ & $\qquad$ & 110 & $\qquad$ & 62\tabularnewline
$ZZ$ &  & $2.1\times10^{4}$ &  & 3 &  & 0\tabularnewline
total &  & $1.3\times10^{5}$ &  & 113 &  & 62\tabularnewline
\hline
\hline 
 &  &  &  &  &  & \tabularnewline
\hline
\hline 
$\begin{array}{c}
{\rm Signal}\\
(m_{H^{0}},m_{A^{0}})\end{array}$  &  & basic &  & optimal &  & $m_{\ell\ell}<10\,{\rm GeV}$\tabularnewline
\hline 
$\left(50,60\right)$ &  & 117 &  & 37 &  & 37\tabularnewline
$S/B$ &  & $9\times10^{-4}$ &  & 0.33 &  & 0.60\tabularnewline
$S/\sqrt{B}$ &  & 0.32 &  & 3.48 &  & 4.70\tabularnewline
\hline 
$\left(50,70\right)$ &  & 433 &  & 56 &  & 50\tabularnewline
$S/B$ &  & $3.3\times10^{-3}$ &  & 0.50 &  & 0.81\tabularnewline
$S/\sqrt{B}$ &  & 1.20 &  & 5.27 &  & 6.35\tabularnewline
\hline 
$\left(50,80\right)$ &  & 680 &  & 38 &  & 26\tabularnewline
$S/B$ &  & $5.2\times10^{-3}$ &  & 0.34 &  & 0.42\tabularnewline
$S/\sqrt{B}$ &  & 1.89 &  & 3.57 &  & 3.3\tabularnewline
\hline
\end{tabular}
\end{table}

We note that a competition between the basic cut and mass window cut
occurs. For example, the acceptance of the basic cut is large but
that of the mass window cut is small for $\Delta m_{A^{0}H^{0}}=30\,{\rm GeV}$.
On the contrary, the acceptance of the basic cut is small but that
of the mass window cut is large for $\Delta m_{A^{0}H^{0}}=10\,{\rm GeV}$.
When the mass gap is less than $5\,{\rm GeV}$, the $p_{T}$ of two
charged leptons is so small that most signal events fail the basic
cut, and hence there is no hope to observe a signal with such a small
mass gap at the LHC.

\newpage{}

\section{Conclusion}

In this work we have shown how the particles of the $dark$ scalar
doublet may be discovered at the LHC, and how their presence will
affect the properties of the SM Higgs boson. From present LEP data,
we obtain the bound $m_{H^{0}}+m_{A^{0}}>m_{Z}$, whereas the SM Higgs
mass bound of 114.4 GeV can be relaxed down to about $106\,{\rm GeV}$,
assuming that $m_{A^{0}}-m_{H^{0}}=10$ GeV. The dark scalars dramatically
affect the search of the SM Higgs boson in the intermediadte mass
region, i.e. $m_{h}\sim100-150\,{\rm GeV}$, at the LHC when the decay
mode $h\to H^{0}H^{0}$ opens. We find that the decay branching ratios
of the usual decay modes of the SM Higgs boson, i.e. $h\to b\bar{b}/WW^{*}/ZZ^{*}/\gamma\gamma$,
are highly suppressed, $\sim60\%$, which makes it more challenging
to observe the SM Higgs boson in those usual decay modes. On the contrary,
one could detect the SM Higgs boson in its decay into the dark scalars,
which will escape the collider detection. We show that it is very
promising to search for the SM Higgs boson through its invisible decay
in the so-called weak-boson fusion process\ \citep{Eboli:2000ze}
which can cover most of the parameter space $\left(m_{H^{0}},\,\mu_{2}\right)$
of the DSDM at the $5\sigma$ significance. After examining the decay
pattern and decay branching ratios of the dark scalars, we consider
their discovery potential at the LHC via the process of $A^{0}H^{0}$
associated production. For $m_{H^{0}}$ of about 50 GeV (which is
also consistent with its being a dark-matter candidate), it should
be observable at the LHC.

\begin{acknowledgments}
This work was supported in part by the U.~S.~Department of Energy
under Grant No.~DE-FG03-94ER40837. GR thanks the Department of Physics
and Astronomy, UCR for hospitality during his summer visit. 
\end{acknowledgments}
\bibliographystyle{apsrev}
\bibliography{reference}

\end{document}